\documentclass[a4paper,11pt]{extarticle}                         
\usepackage{extsizes} 
\usepackage[english]{babel}
\usepackage{graphicx,color}
\usepackage[utf8]{inputenc}
\usepackage{amsmath,amssymb,amsfonts}
\usepackage{graphicx}
\usepackage[hmarginratio=1:1,top=32mm,columnsep=20pt]{geometry} 
\usepackage{multicol}                                            
\usepackage[small]{caption}         
\usepackage[sc]{mathpazo}         
\usepackage{booktabs}
\usepackage{float}
\title{\vspace{-15mm}{\huge Optimizing Beer Glass Shapes to Minimize \\  Heat Transfer During  consumption} }
	\author{\large\it{Cláudio C. Pellegrini}	\\ 
	\small  Federal Univ. of São João del-Rei \\ \small Depart. of fluid and thermal sciences, São João del-Rei, MG, Brazil.  \\ 
	\small Corresponding author: pelle@ufsj.edu.br
}
\date{}

\begin{document}
\maketitle  

% ###################################################
\begin{abstract}
	\noindent
	This paper addresses the  problem of determining  the optimum shape for a beer glass that minimizes the heat transfer while the liquid is consumed, thereby keeping it cold for as long as possible. The  proposed solution avoids the use of insulating materials. The glass is modelled as a  body of revolution generated by a smooth curve $S$, constructed from a material with negligible thermal resistance at the revolution surface but insulated at the bottom.  The ordinary differential equation describing the problem is derived from the first law of Thermodynamics  applied to a control volume encompassing  the liquid. This is an inverse optimization problem,  aiming to find the shape of the glass (represented by curve $S$)  that minimizes the heat transfer rate. In contrast, the direct problem  aims to determine the heat transfer rate for a given geometry.  The solution obtained is analytic, and the resulting expression  for $S$ is in closed form, providing  a family of optimal glass shapes that can be manufactured using conventional methods.
		
\end{abstract}

% =================================================================== 
\begin{multicols}{2}    
\section{Introduction}

An effective pedagogical strategy in mathematical and physics education involves demonstrating to students how the theories taught in the classroom can be used to address everyday problems.
%A valuable approach in the teaching of physics, is to show students how  everyday problems can be dealt with by the  theories presented in  classroom. % 
Nevertheless, this proves to be a difficult task more often than not. 

Few everyday problems  in physics possess a level of simplicity  that  allows for analytical treatment, while still capturing  all  phenomena involved. Introducing too many simplifications may render the problem  unrealistic\footnote{For example, consider an   airplane  wing represented by a perfectly smooth, infinitely thin  flat plate inclined by an angle $\theta$ in relation  to a  frictionsless airflow.}. If too few simplifications are emploied, the problem may become unsuitable for analytical treatment\footnote{The same airplane wing now is subjected to  turbulent airflow and presents a rough surface defined by given functions $y_{up}(x)$ and $y_{low}(x)$, where $x$ is the position along the chord  and $y_{up}$ and  $y_{low}$ are respectively the upper and lower half thickness at a given value of the position $x$}. 

In the field of mathematics, finding practical  problems that are easy to understand but can be solved by straightforward techniques, is an even more challenging task. 
The four-color theorem\footnote{Given any separation of a plane into contiguous regions, producing a figure called a map, no more than four colors are needed to color the regions of the map, so that no two adjacent regions have the same color.} (\cite{appel1} and \cite{appel2}) and the hairy ball theorem\footnote{You can't comb a hairy ball without creating a whirlpool.} \cite{eisenberg} are traditional examples. Both require considerable expertise to  understand the solution. More recently, ideas about how to divide a pizza into an arbitrary number of unconventionally divided parts   \cite{Humenberger}  have also gained interest, but is equally challenging to understand.

Motivated by the last problem, the authors envisioned an analysis related to the beverage that arguably (or not) pairs best with pizza: beer. In tropical countries like Brazil, a recurring problem is how to keep it cold during consumption, especially in coastal regions.
 
This paper considers the  problem of finding the optimum shape of a beer glass such that the heat transfer rate is minimized in order to keep the beverage cold for as long as possible, {\sl while it is consumed}. This is an inverse optimization problem,  in the sense that the objective is to find the geometry that minimizes the heat transfer rate as opposed to the direct problem, where the objective would be to find the heat transfer rate associated with a given geometry. 
The intention is to encourage engineering students to develop a rational approach to problems in physics, abandoning the fairly common concept that "theory in practice is different". As a final contribution, nevertheless, the study also aspires to help improve  the palatability of our beers.
%The present research proposal does not constitute an apology for alcohol consumption, emphasizing the importance of always drinking in moderation.

A brief search on the literature shows literally  hundreds of articles focused on analysing practical problems across diverse areas of physics, but nothing was found on heat transfer, except for  \cite{Planinsic}, which  addresses the heating of cheese. 

The problem of keeping a liquid  in a reservoir at the lowest possible temperature may be solved, as will be shown later, by finding a surface that minimizes the area-to-volume ratio of the reservoir. The Greeks  knew, even though they could not prove this rigorously, that  the answer to the two-dimensional  version of this  problem was the circle. Later findings showed that the solid possessing the lowest surface-to-volume ratio was the sphere. However, formal proofs for both cases had to wait until the 19th century. In contemporary times, this problem is addressed through the application of the isoperimetric inequality\footnote{Let $\gamma$ be a closed, piecewise differentiable plane curve of class $C^1$. Let $L$ and $A$ be its perimeter and surface, respectively. The isoperimetric inequality establishes that $4\pi A \leq L^2$, with the equality valid if $\gamma$ is a circle.} in three dimensions (for which there is a variety of proofs) or through variational calculus concepts.  
 
Applications of surface-to-volume ratio optimization are numerous and extend to practically all areas. In chemistry, there are studies involving reactions of all types, such as combustion (in engines or fires), drying and humidification of particles. In biology, there are studies involving exchanges through the skin of living beings and the membrane of cells, microorganisms or organelles. In engineering, there are studies of heat transfer in reservoirs and heat and mass transfer in systems subject to phase change. In atmospheric sciences, there are studies involving the formation of raindrops, hail and snowflakes, as well as evaporation in vegetables and bodies of water. In pharmacology, there are studies on the absorption of medications. It is unnecessary to point out the existence of a series of multidisciplinary studies, with an interface between different areas of study. 
%The study by \cite{Planinsisic}, for example, shows how the concept of maximizing the surface-to-volume ratio can be applied to heating cubic pieces of cheese and then shows its relationship with animal metabolism, taking the example of rats, the proverbial cheese eaters. 

A common factor in all the  above mentioned studies, is the fact that the heat exchange surface does not change shape during the process. 
This, however, is not the case when considering the heat exchanged by a glass of liquid with the surroundings during consumption. Even in  a cylindrical glass, the total exchange surface undergoes changes in shape during consumption. As the liquid level lowers, the side area reduces while the upper area remains preserved. 

Therefore, the question we intend to answer here is: what is the optimal shape-varying surface that minimizes the heat transfer by convection on a glass of liquid with its surroundings? In other words, what is the ideal  beer glass, considering potential variations in shape as the liquid is consumed?

% =================================================================== 
\section{Mathematical model}
The process is quite straightforward here: a request is made for a beer, the waiter delivers it, it is served,   it is consumed. Repeat. 

Once poured in the glass, the beer begins to exchange heat with its surroundings, a process that lasts until it attains thermal equilibrium with of the environment (including the glass), a result that essentially  nobody wants. Depending on the initial temperature difference between the beer and the surroundings, within a short period of time the drink  may become unsuitable for consumption. In the most favourable  case, the environment  at 10 \textdegree C and the beer\footnote{Who drinks beer below 10 \textdegree C?}   at 4 \textdegree C, personal experience shows that as long  as 30 minutes may pass before the beer gets warm. In the most critical scenario, such as at the beach on a 38 \textdegree C  windy day, as few as 3 minutes  may be sufficient (again based on personal experience, exhaustively repeated) to render the beer undrinkable. 

There are several established practical methods to  decrease  the beer's heat transfer with the environment. The use of insulation tubes made of expanded polystyrene (EPS or Styrofoam${}^{\text \textregistered}$) is probably the most common and it is also used for beer bottles in Brazil. The use of handles on mugs is also a common method to isolate the  consumer's  hand heat from  the drink inside the container. The habit of maintaining a layer of foam on the top of the beer  acts as a thermal insulator due to its low conductivity. In addition, it also prevents excessive loss of CO${_2}$. All those methods have in common the fact that they are  {\sl passive}, i.e., they do not depend on any heat transfer device or substance. That is exactly the approach we shall use here. 

\begin{figure} [H]
   \centering \includegraphics[trim={50mm 0 50mm 0},clip,width=0.9\columnwidth]{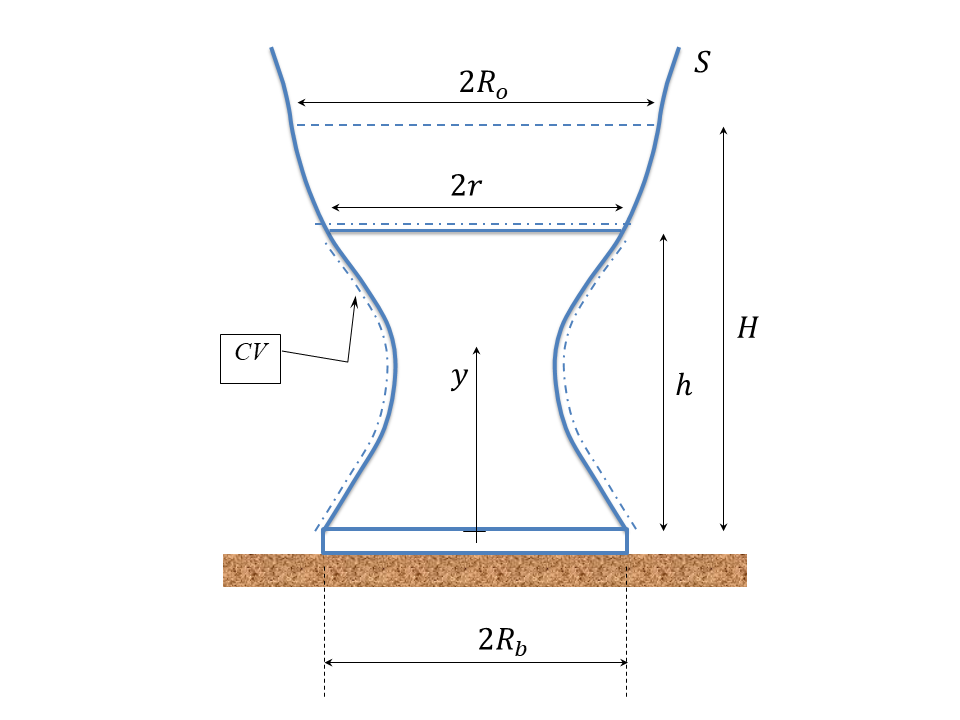}
   \caption{Typical portable liquid container device.}
   \label{fig:copo}
\end{figure}

To state the most general version of the problem, consider a container  holding a  liquid initially not in thermal equilibrium with its surroundings, as shown in Fig \ref{fig:copo}. Assume that the vessel is a body of revolution, generated by the rotation of a curve $S$, differentiable of class $C^1$, around the vertical axis, $y$. Such a  geometry  describes all commercially available containers, except for  some non-axisymmetric non-returnable beer bottles. However, $S$ is not entirely unrestricted: it must contain exactly one opening and one impermeable bottom.
 
Therefore, let $0 \leq h(t) \leq H$ be the variable height of the liquid in the container  at time $t$, and let $0\leq r(t)$ be the corresponding radius, where $r(0)=R_b$ is the  radius of the  base and $r(H)=R_o$ is the radius of  the opening. For obvious reasons, $r=0$ may only occur at $h=0$. Let CV be a variable shape control volume  encompassing  the liquid,  as depicted in Fig.~\ref{fig:copo}. 

To begin particularising the problem, now assume  that body of the vessel has negligible thermal resistance, whereas its bottom is thermally insulated. This is a reasonable approximation for  beer glasses, where the body is made of thin glass for comfort, and the bottom  of   thick glass for mechanical resistance. Also contributes for making the bottom insulation a reasonable hypothesis the habit of letting beer glasses rest on insulating surfaces, such as  tables,  counters and cardboard disks (the famous "wafer").
 
Furthermore, consider that the temperature of the beer remains uniform throughout the glass while it is being consumed. This holds true when the rate of temperature change over time is small, enabling the liquid to  reach instantaneous thermal equilibrium. Under such circumstances, the initial temperature difference between the liquid and the environment must not be excessively large, which is always the case.

Finally, assume  the liquid to be  homogeneous. This characteristic holds true for most filtered beers, that do not form accumulations at the bottom, including {\sl Weizenbier}. A few  craft beers  present small amounts of yeast sediment and are not considered here.

The law of conservation of energy (\cite{incropera} for example) for the  CV chosen is  
\begin{equation}  \label{eq:plt02}
q_{ac} = q_{in} - q_{out} + q_{gen}
\end{equation}
where $q=dQ/dt$ is the heat transfer rate and $Q$ is heat. 

In Eqn. \eqref{eq:plt02}, $q_{ac}$ represents the  accumulated heat within the CV, $q_{gen}$ is the heat generated inside the CV, and $q_{in}$ and $q_{out}$ denote the heat entering and leaving the CV, respectively.  

In general, while drinking beer and most beverages, there is no internal heat generation. Possible exceptions would be liquids undergoing fermentation. Thus, in our case, $q_{gen}=0$. Assuming that the  environment is the only external source of heat, Eqn. \eqref{eq:plt02} yields
\begin{equation}  \label{eq:plt2}
q_{ac} =  q_{in}
\end{equation}
where $ q_{in}$ enters through the body of the glass and the foam at the opening. 

Using the concept of thermal resistance, Eqn.~\eqref{eq:plt2} can be rewritten as
\begin{equation}  \label{eq:plt3}
q_{ac} = \frac{T - T_\infty}{ R_\text{side}} +  \frac{T - T_\infty}{R_\text{foam}}
\end{equation}
where $T=T(t)$ is the spatially uniform temperature of the beer, $T_\infty$ is the ambient temperature, $R_\text{side}$ and $R_\text{foam}$ are  the thermal resistances  of the glass and of the foam respectively. Each resistance is composed by two other resistances associated in series: a conductive resistance  through the materials and a convective resistance on  their external surfaces. However, to simplify the problem, the thermal conductivity of the side glass and the foam are neglected\footnote{Both depend on the shape of the surface, which has not yet been determined, rendering the solution iterative. }. This may not be completely realistic but nevertheless represents the most critical situation\footnote{The absence of foam can also be considered, according to the experts,  a tasting heresy}.
Equation \eqref{eq:plt3} may then be  rewritten as
\begin{equation}  \label{eq:TRC1}
	q_{ac} =  \frac{T - T_\infty} {R_\text{side}^\text{cv}}
	+ 	\frac{T - T_\infty} {R_\text{foam}^\text{cv}}
\end{equation}

From the definitions of specific heat and density for an homogenous liquid it follows that $q_{ac} = \rho V c_p ({dT}/{dt})$. Thence,
\begin{equation}  \label{eq:TRC2}
\rho V c_{p} \frac{dT}{dt} =  \frac{T - T_\infty} {R_\text{side}^\text{cv}} 
+
\frac{T - T_\infty} {R_\text{foam}^\text{cv}} 
\end{equation}

Substituting $R^\text{cv} = 1/(h_\text{cv}A)$ finally yields 
\begin{equation}  \label{eq:TRC3}
\rho V c_{p} \frac{dT}{dt} =  h_\text{cv} A_\text{tot}(T - T_\infty)
\end{equation}
where $A_\text{tot}$ is the total heat transfer  area, i.e., the side and opening areas of the glass.

%Under  the hypothesis used, Eqn. \eqref{eq:TRC3} shows that the temperature variation results only from the convective heat transfer through the exposed area of the glass and the foam. Therefore, the shape of the glass is the only element responsible for  the temperature variation.

Equation \eqref{eq:TRC3} suggests some strategies for minimizing the rate of temperature change, $dT/dt$, thereby extending the drinkability window:
\begin{enumerate}
	\item Reduce $(T - T_\infty)$ by keeping the glass in a cool environment and avoiding radiant heat from the sun\footnote{Which was not considered in the present mathematical modelling and would represent an extra term in Eqn. \eqref{eq:plt3}.};
	\item Increase the sum of resistances in the denominator of Eqn. \eqref{eq:TRC3} by keeping a tick, generous foam over the beer;
	\item Increase the  conductivity resistance of the sides of the glass (the vessel) by substituting the glass (the material) by a more insulating material, such as ceramic\footnote{This strategy is often used  on large mugs, alongside with a handle to keep the consumer's hand out of contact with the side of the glass, guaranteeing no other external heat source than the environment. Unfortunately, it makes lip contact uncomfortable, some say.};
	\item Keep the glass  away from drafts, avoiding forced convection, which is far more efficient than natural convection in transfering heat.
\end{enumerate}

The considerations above  show why the beach is the most challenging environment for beer drinking: the air temperature is high,  the wind is persistent, the sun shines, and ceramic mugs are not welcome. 

Rearranging Eqn. \eqref{eq:TRC3} yields
\begin{equation}  \label{eq:TRC4}
 \frac{dT}{dt} = \frac{ h_\text{cv}}{\rho c_{p}} \left(\frac{A_\text{tot}}{V}\right)(T - T_\infty)
\end{equation}

%Supondo, sem perda de generalidade, um determinado instante fixo do tempo, a equação \eqref{eq:TRC4} mostra que a taxa de variação instantânea da temperatura do líquido, $dT/dt$, é tão menor quanto menor for a relação ${A_\text{tot}}/{V}$ para um dado valor fixo de $h_\text{cv}/ (\rho c_{p})$. Deste modo, minimizar a heat transfer reduz-se a minimizar a relação ${A_\text{tot}}/{V}$, assunto que será tratado na próxima seção. 
Equation \eqref{eq:TRC4} is the ODE that governs the problem. It can be easily  solved for a known geometry, since it is of the separable type. %Rewriting it as $d\Theta/\Theta =Bdt$ where $\Theta=T - T_\infty$ and $B = h_\text{cv} A_\text{tot}/ (\rho V c_{p})$, a solution that tends exponentially to $\Theta = 0$ is  obtained. 
However, this would be the direct problem which is not the goal here.

For any fixed instant of time $t$ and any given value of $h_\text{cv}/ (\rho c_{p})$, Eqn. \eqref{eq:TRC4} shows that $dT/dt$ gets smaller as  $ {A_\text{tot}}/{V}$ is reduced. Therefore, minimizing the heat transfer  reduces to 
\begin{equation}
	\text{minimize}\left(R=\frac{A_\text{tot}}{V}\right),\quad t\in \mathbb{R}^+
\end{equation}

This is the subject of the next section.

% =================================================================== 
\section{Solution}
\begin{figure} [H]
   \centering \includegraphics[trim={10mm 0 20mm 0},clip,width=0.9\columnwidth] {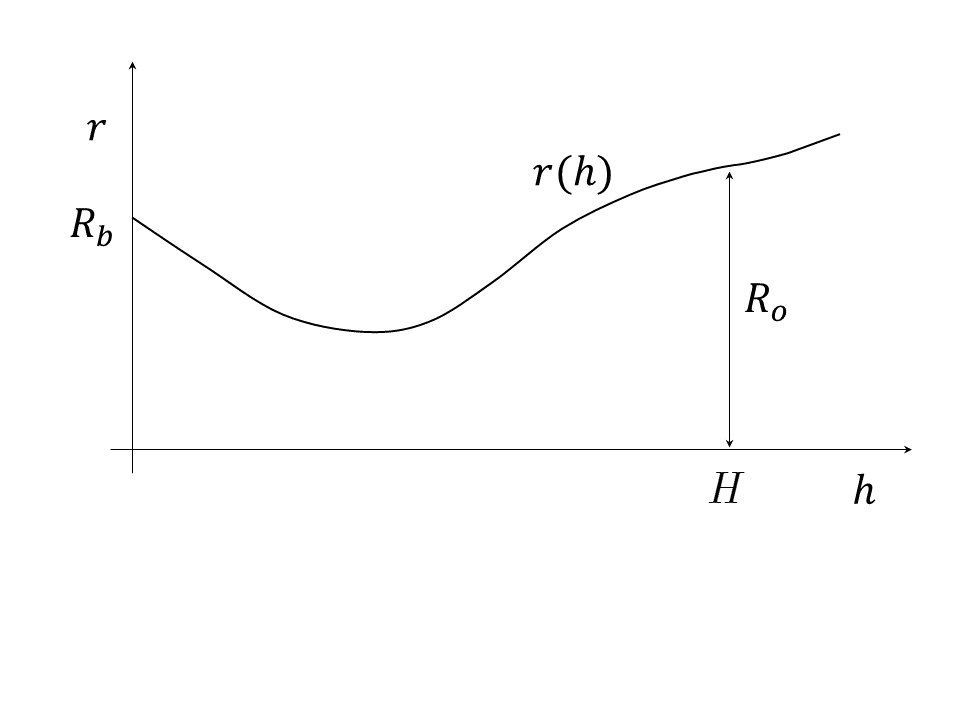}
   \caption{Glasses' generating curve}
   \label{fig:copo1}
\end{figure}

Consider Fig. \ref{fig:copo1}, where the glass and its generating curve, here identified as $r=r(h)$, are depicted
% and $r \geq 0$ , $r(0)=R$, $r(H)=R_o$ and $0 \leq h \leq H$. The presumed continuity and differentiability of $r(h)$ suits both the mathematical analysis and the glass manufacturing process. The radius $R_o$ refers to the limit of the liquid, which may not coincide with that of the glass. 

The  minimum for the surface-to-volume ratio follows from 
\begin{equation}  \label{eq:minimo1}
\frac{d}{dh} \left(\frac{A_\text{tot}}{V}\right) = 0
\end{equation}

Here, the height of the liquid, $h$, is used as the independent variable as  the beer is consumed. Differentiation of Eqn. \eqref{eq:minimo1} yields
\begin{equation}  \label{eq:minimo1.5}
\frac{A'_\text{tot}V - A_\text{tot}V'} {V^2}=0
\end{equation}

Simplifying  results in
\begin{equation}  \label{eq:minimo2}
\frac{A'_\text{tot}}{A_\text{tot}} =\frac{V'}{V}
\end{equation}
for  $V \neq 0$ and $A_\text{tot}\neq 0$. 
Integration gives $\ln{A_\text{tot}} =\ln{ V} + C_0$ and, thus,
\begin{equation}  \label{eq:minimo3}
A_\text{tot}=C_1 V
\end{equation}
where $C_1$ is a dimensional constant  to be examined later.

From Calculus, for bodies of revolution the side area is
\begin{equation}  \label{eq:area}
A_\text{side} = 2\pi \int_0^h{r\sqrt{1+r'^2}\,dh} 
\end{equation}
and the volume is
\begin{equation}  \label{eq:volume}
V = \pi  \int_0^h r^2\,dh
\end{equation}

Therefore, Eqn. \eqref{eq:minimo3} becomes
\begin{equation}  \label{eq:minimo4}
\pi r^2 + 2\pi \int_0^h{r\sqrt{1+r'^2}\,dh} = C_1 \pi  \int_0^h r^2\,dh
\end{equation}

Differentiating in relation to $h$ results in\footnote{Note that the derivative of an integral is the integrand itself as long as the upper integration limit coincides with the differentiated variable, as is the case here.}
\begin{equation} \label{eq:minimo4.3}
	2\pi r r' + 2\pi r \sqrt{1+r'^2} = 2C_2 \pi r^2
\end{equation}
where  $C_2 = C_1/2$. 

Simplifying and rearranging,
\begin{equation} \label{eq:minimum5}
	\sqrt{1+r'^2} =C_2 r - r'
\end{equation}
for $r \neq 0$, except possibly at $h=0$. 
Squaring both sides and simplifying,
\begin{equation} \label{eq:edo1}
	1 = C_2^2 r^2 - 2C_2rr'
\end{equation}

This is a separable ODE. Thus
\begin{equation} \label{eq:edo2}
	\int_0^h{dh} = \int_R^r{ \frac{ 2C_2r}{C_2^2 r^2 - 1}dr}
\end{equation}
which can be integrated by substitution of $\eta = C_2^2 r^2 - 1$. The result is
\begin{equation} \label{eq:h1}
	h = \frac{1}{C_2}\ln ({C_2^2 r^2 - 1}) - \frac{1}{C_2}\ln ({C_2^2 R^2 - 1})
\end{equation}

Recalling that $C_1 = 2C_2$ and rearranging yields
\begin{equation} \label{eq:h2}
	h = \frac{2}{C_1} \ln \left(\frac{ C_1^2 r^2 - 4} {C_1^2 R^2 - 4}\right)
\end{equation}
which can be rewritten to make $r$ explicit as
\begin{equation} \label{eq:h}
	r = \frac{1}{C_1}\sqrt{4 + (C_1^2 R^2 - 4)e^{C_1 h/2}}
\end{equation}

This equation solves the problem.
\smallskip

Substituting Eqn. \eqref{eq:h} into Eqns. \eqref{eq:area} and \eqref{eq:volume} and integrating  from 0 to $H$ yields the total area and volume of the glass:
\begin{equation} \label{eq:area3}
	A_\text{tot} = \frac{\pi}{C_1^2}\left(4C_1 h + 2(C_1^2 R^2-4) e^{C_1 h/2}\right)
\end{equation}
\begin{equation} \label{eq:volume3}
	V = \frac{\pi}{C_1^3}\left(4C_1 h + 2(C_1^2 R^2-4) e^{C_1 h/2}\right)
\end{equation}

\section{Discussion}
Equation \eqref{eq:h} shows an exponential dependence of $r$ with $\sqrt{h}$. One should therefore expect a  rapid growth of $r$ with $h$, affected by the parameter $C_1$ which cannot be chosen arbitrarily, as demonstrated subsequently. 

From Eqn.  \eqref{eq:h2}, 
\begin{equation} \label{eq:restricao2a1}
	\frac{ C_1^2 r^2 - 4}{ C_1^2 R^2 - 4}>0
\end{equation}	
 and therefore 
\begin{equation} \label{eq:restricao2a}
	C_1 r > 2
\end{equation}
for $R_b \leq r \leq R_o$. Remember that from Eqn. \eqref{eq:minimo3},  $C_1>0$, for $0<r\leq R$. 

%As $C_1 > 0$, negative intervals are excluded and it simply results
%\begin{equation} \label{eq:restriction}
%	C_1 r \neq 2 \quad \text{e} \quad C_1 > 0
%\end{equation}
%for all $r$, including $r=R$.

Equation \eqref{eq:h} also shows that the glass must have a small base and a large opening, with its radius continually  increasing from the base up.  Indeed, putting  $r'=0$ in Eqn. \eqref{eq:edo1}  implies  that $C_1 r =2$, which is  false. Therefore, that is no maximum or minimum points in  function $r$. Additionally, from Eqn. \eqref{eq:edo1} and $C_1=2 C_2$ it follows that $r'= (1 - C_1^2 r^2/4)/C_1 r$ which is always positive, showing that $r(h) $ is a monotonically increasing function.

Before  Eqn. \eqref{eq:h} is sent to the factory, and the optimized glasses start to be mass produced, it is necessary to establish adequate values for $C_1$, limited by the restriction $C_1 R>2$ as $C_1$ is a free parameter in Eqn. \eqref{eq:h2}. 
 
In Eqn. \eqref{eq:h2}, as  $C_1 R \rightarrow 2$ from the right for fixed values of $r$, then $h \rightarrow \infty$ for any finite $C_1$. In words, as $C_1 R \rightarrow 2$ it is only possible to obtain the optimal solution for very tall glasses. This should not be an issue as one could simply chose  a large value of $C_1 R$ to avoid the problem. However, as $C_1 R$ multiplies $\exp(C_1h/2)$, large values of $C_1 R$ will result in very large values of $r$. Thus as $C_1 R \gg 2$ it is only possible to obtain the optimal solution for very large openings. Both situations suggest that there exists a suitable range of values of $C_1 R$ that results in practically viable glass shapes. 

To obtain $C_1$ first Eqn. \eqref{eq:h2} is rewritten as
\begin{equation} \label{eq:h3}
	e^{(C_1h/2)} = \frac{ C_1^2 r^2 - 4} {C_1^2 R_b^2 - 4}
\end{equation}

Then, based on practical considerations, values of $R_o$  and $\lambda = R_o / R_b$ must be chosen. Substituting   $r=R_o = \lambda R_b$ and $h=H$ into Eqn.~\eqref{eq:h3} then yields
\begin{equation} \label{eq:h4}
	e^{(C_1H/2)} = \frac{ (\lambda C_1 R_b)^2 - 4}{(C_1 R_b)^2 - 4}
\end{equation}

Supposing that $ {C_1 R_b} \approx 2 $ to avoid very tall glasses implies that $(\lambda C_1 R_b)^2 - 4 \approx 4(\lambda^2 -1)$ and that $ {(C_1 R_b)^2 - 4} = \varepsilon \approx 0$. 
Equation \eqref{eq:h4} then becomes 
\begin{equation} \label{eq:eps}
	{ \varepsilon} \approx { 4(\lambda^2 -1)}e^{-C_1H/2}
\end{equation}

Therefore, $ \varepsilon = {C_1^2 R_b^2 - 4} $ can be re\-writ\-ten as
\begin{equation} \label{eq:eps2}
	C_1R_b = \sqrt{ \varepsilon + 4}
\end{equation}

Equations \eqref{eq:eps} and \eqref{eq:eps2} 
provide an approximate way to obtain $C_1$:
\begin{enumerate}
	\item Values of $R_o $ and $R_b $ are chosen;
	\item An initial approximation for $C_1$ is obtained through $C_1R_b=2$. This value is not final because it makes  the denominator of Eqn. \eqref{eq:eps2} go to zero;
	\item This value is then substituted into Eqn. \eqref{eq:eps}  to calculate $\varepsilon$;
	\item This  value obtained for $\varepsilon$ is substituted into Eqn. \eqref{eq:eps2} to give a final estimate for $C_1$
\end{enumerate}

As $C_1$ is used only as a shape parameter, there is no need for further iterations or a convergence criterion.

\begin{figure} [H]
   \centering \includegraphics [trim={0mm 0 0mm 0},clip,width=0.9\columnwidth] {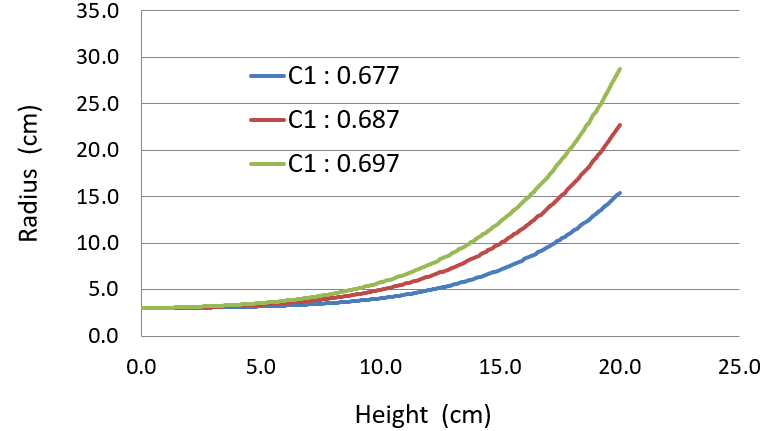}
   \caption{Optimized glass shapes for $R_b=3.0$ cm.}
   \label{fig:funcao1}
\end{figure}

\begin{figure} [H]
   \centering \includegraphics [trim={0mm 0 0mm 0},clip,width=0.9\columnwidth]{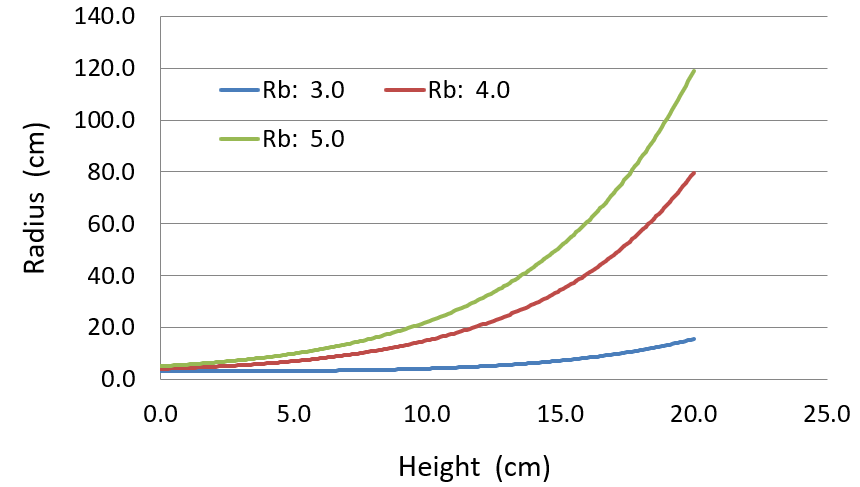}
   \caption{Glass shapes for $C_1=0,677$ cm$^{-1}$.}
   \label{fig:funcao1b}
\end{figure}

The  influence of $C_1$ and $R_b$ on the shape of the glass is illustrated in Figs. \ref{fig:funcao1} and \ref{fig:funcao1b}, for glasses with $H=20.0$ cm. The numerical results are presented in Tables  \ref{tab:glasses1} and \ref{tab:glasses2}.
	
Figure \ref{fig:funcao1} shows the results obtained for three glasses with a    base radius $R_b=3.0$ cm and  $C_1$ varying around 0.667 cm\textsuperscript{-1}, the value calculated by  Eqn. \eqref{eq:eps} with $\lambda=2$. Only  glass 1, with $C_1=0.677$ cm\textsuperscript{-1} is commercially acceptable. Glasses 2 and 3  are certain to have  equilibrium issues due to their large relation $R_o/R_b$. Additionally, their volumes are so large that it is very unlikely  the beer final temperature is acceptable, despite the fact they are thermally optimized\footnote{No, you don´t know somebody that drinks {\sl that} fast.}.  
\begin{table}[H] 
	\caption{Optimized glass shapes. $H=20.0$ cm and $R_b=3.0$ cm }
	\centering
	\begin{tabular}{c|c|c|c|c}
		\toprule
		N &  $C_1$  & $R_o$   & $V_{tot}$  & $R_o/R_b$\\
		 &  (cm${}^{-1}$) &  (cm) &  (litres) & \\
		\midrule
		1 & 0.677 & 15.4 & 2.68 & 5.1 \\
		2 &	 0.687 & 22.8 & 5.15 & 7.6 \\
		3 &   0.697 & 28.7 & 7.87 & 9.6 \\
		\bottomrule
	\end{tabular}
	\label{tab:glasses1}
\end{table}

Figure \ref{fig:funcao1b} and Table \ref{tab:glasses2} shows the results obtained for three glasses with  $C_1=0.677$cm\textsuperscript{-1} and $R_b$  varying. Again, only  glass 3  is acceptable with a volume of 7.87 litres. As seen on Tab. 	\ref{tab:glasses2} the volumes of  glasses 4 and 5 are absolutely ridiculous. 
%To keep $C_1 R > 2$ with  reasonable base radii, say between 1 and 5 cm, we have $C_1$ between 2/5 and 2 cm$^{-1}$. This implies an acceptable increase in radius for $C_1 \approx 2/5$ but unacceptable for $C_1 \approx 2$, resulting in $R_o = 101$ m at 20.0 cm height.
\begin{table}[H] 
	\caption{Optimized glass shapes. $H=20.0$ cm and $C_1=0.677$ cm\textsuperscript{-1} }
	\centering
	\begin{tabular}{c|c|c|c|c}
		\toprule
		N &  $R_b$  & $R_o$   & $V_{tot}$  & $R_o/R_b$ \\
		 &  (cm${}^{-1}$) & (cm) &  (litres) & \\
		\midrule
		3 & 3.0 & 15.4 & 2.68 & 5.1 \\
		4 &	4.0 & 79.7 & 59.36 & 19.9 \\
		5 & 5.0 & 119.1 & 132.14 & 26.4 \\
		\bottomrule
	\end{tabular}
	\label{tab:glasses2}
\end{table}

\begin{figure} [H]
   \centering \includegraphics [trim={0mm 0 0mm 0},clip,width=0.9\columnwidth]{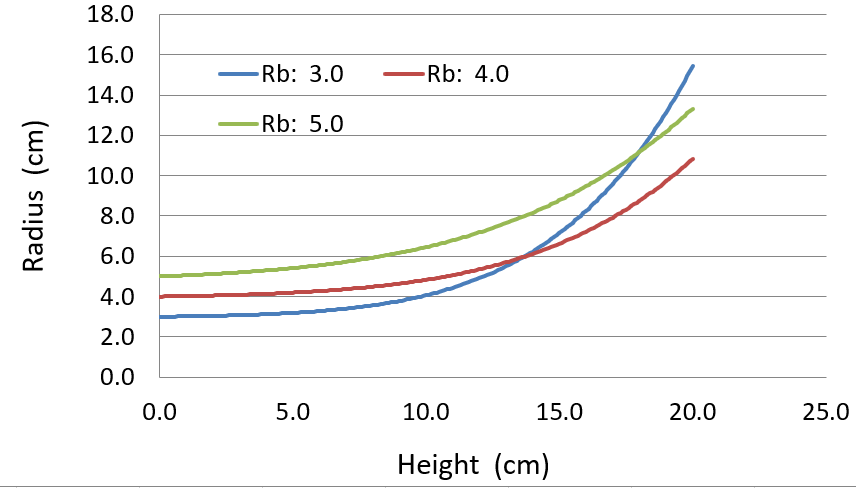}
   \caption{Glasses 3,6 and 7}
   \label{fig:funcao2}
\end{figure}

Due to the simplicity of the calculations required for obtaining optimized glasses, many alternative configurations were explored. Of all resulting glasses, two were considered feasible. Their dimensions are provided in Table \ref{tab:glasses3}, alongside those of glass 3. Figure \ref{fig:funcao2} shows a comparison between  these three  glasses, numbered 3, 6 and 7. %7, showing that an excessively cylindrical shape was avoided. 
\begin{table}[H] 
	\caption{Optimized glass shapes. $H=20.0$ cm }
	\centering
	\begin{tabular}{c|c|c|c|c}
		\toprule
		N &  $R_b$  & $C_1$   & $R_o$  & $V_{tot}$ \\		
		  &   (cm) &  (cm${}^{-1}$)  &  (cm) & (litres) \\
		\midrule
		3 & 3.0 & 0.677 & 15.4 & 2.68 \\
		6 &	4.0 & 0.510 & 10.8 & 2.22 \\
		7 & 5.0 & 0.420 & 13.3 & 3.74 \\
		\bottomrule
	\end{tabular}
	\label{tab:glasses3}
\end{table}

\begin{figure} [H]
   \centering \includegraphics [trim={10mm 0 0mm 0},clip,width=0.9\columnwidth] {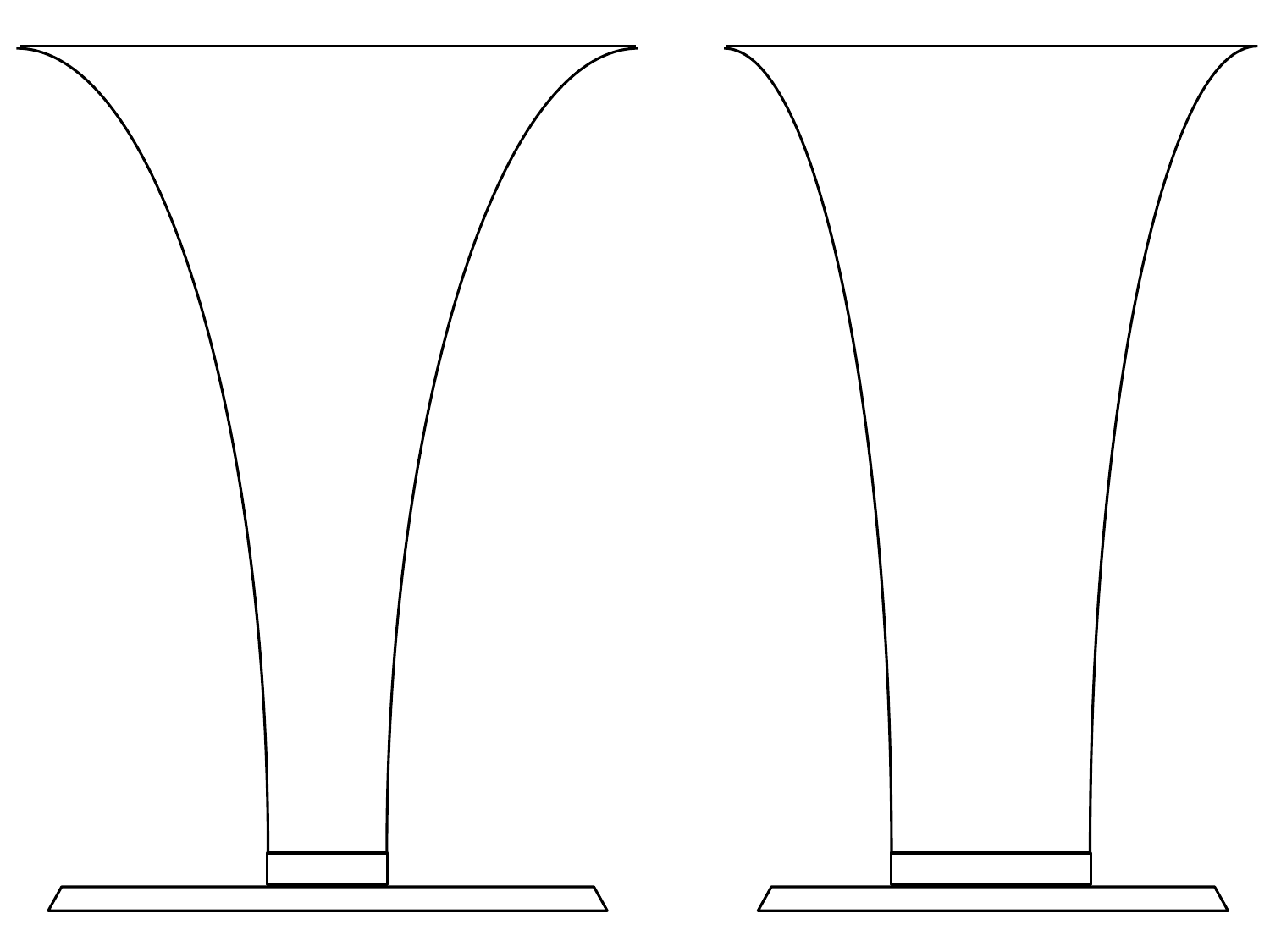}
   \caption{Glasses 3 and 7 in scale}
   \label{fig:copo4}
\end{figure}

\begin{figure} [H]
   \centering \includegraphics [trim={10mm 0 0mm 0},clip,width=0.5\columnwidth] {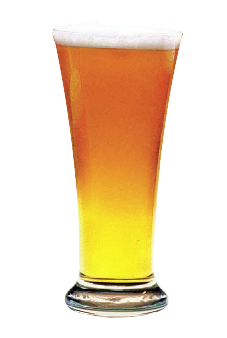}
   \caption{Over-the-shelf glass similar to glass~3.}
   \label{fig:copo5}
\end{figure}

All of them have total volumes  considered high by most non-alcoholic consumers and considerably large radius of the opening, requiring a wide foot to keep balance. Smaller radius of the base would make $C_1$  increase, leading to even wider openings. Larger radius of the base, on the other hand, would yield small values of $C_1$, resulting in almost cylindrical bodies, which would require very tall glasses to be optimized. Figure~\ref{fig:copo4} presents an in scale shape comparison between glasses 3 and 7. An image of a commercial glass similar to glass 3 is presented  in figure \ref{fig:copo5}.

Here, as in the case of cutting pizzas, the mathematically optimized solution to the problem may prove to be somewhat complicated to implement in practice.

\section{Conclusions}
In this work, a method has been proposed to optimize the shape of  beer containing glasses. The optimizations criterion was to minimize heat transfer, thus maintaining a low temperature for as long as possible while the liquid is  consumed. The analysis yielded a family of shapes that  can be easily manufactured by traditional methods and used in day-to-day life, as long as some attention is pay to the  radii of the base and of the opening.

Throughout the analysis several hypothesis were used. The glass was considered to be a body of revolution generated by the rotation of a smooth curve $S$ around the vertical axis. Additionally, the  thermal resistance on the body of the glass was neglected, whereas the bottom was considered insulated. The liquid's temperature was supposed to be spatially uniform, the liquid was considered to be  homogeneous and the thermal resistance of the foam was disregarded.  Finally, no radiative  heat transfer was considered, as well as the conduction heat transfer due to hand contact with the body of the glass. 

While those hypotheses may appear restrictive,  the analytical result obtained still hold didactic and practical interest despite the fact that a complete  formulation including all effects disregarded earlier, can be effectively addressed  by numerical methods. Nevertheless, given that the primary focus of the investigation conducted here is  didactic, analytical solutions are preferable.  Closed analytical solutions  are generally  welcomed in Physics, even though they usually represent the result of simplified analyses. They offer a comprehensive  view of the problem, explicitly depicting  the influence of all parameters involved. 

Moreover, an analytical solution is almost always a general conclusion about the problem addressed and rarely constitutes a case study. Analytical solutions also clarify the conditions under which the obtained results   are valid. These statements may seem somewhat obvious, but in times of such intense and careless use of computer simulation, they are particularly timely.

Potential areas of further investigation include allowing the base of the glass and its body to exchange heat with the ambient, including radiative heat transfer and/or conduction  heat transfer through the foam. Some of these analysis may require differential methods volume or an iterative procedure. 

In conclusion,  this paper applied basic concepts of  heat transfer and  extreme values of functions to an everyday, yet relevant topic -- beer drinking. The primary goal, of course, was to enhance  engineering students' interest in Physics and Mathematics. However, a secondary yet  crucial application of the present results was to safeguard the quality of our beers.

% =================================================================== 

\end{multicols}

\begin{thebibliography}{9}
\bibitem{appel1} Appel, K.; Haken, W., Every Planar Map is Four Colorable. I. Discharging, Illinois Journal of Mathematics, 21 (3): 429–490, doi:10.1215/ijm/1256049011, 1977
\bibitem{appel2} Appel, K.; Haken, W.,; Koch, J., Every Planar Map is Four Colorable. II. Reducibility, Illinois Journal of Mathematics, 21 (3): 491–567, doi:10.1215/ijm/1256049012, 1977
\bibitem{eisenberg} Eisenberg, Murray; Guy, Robert (1979), A Proof of the Hairy Ball Theorem, The American Mathematical Monthly, 86 (7): 571–574, doi:10.2307/2320587
\bibitem{Humenberger}  Humenberger, H. Dividing a pizza into equal parts –- an easy job? The Mathematics Enthusiast, (12) 1, p. 389-403, 2015.
\bibitem{Planinsic} {G. Planinsic, M. Vollmer}, European Journal of Physics, \textbf{29}, 369-384 (2008).
\bibitem{incropera} {F. P. Incropera, D. P. DeWitt},  Fundamentals of heat and mass transfer. John Willey  \& Sons, 6th ed. ISBN-13 978-0471457282, 2006


\end{thebibliography}
\end{document}